# Percolation induced gel–gel phase separation in a dilute polymer network

Shohei Ishikawa[1], Yasuhide Iwanaga[1], Takashi Uneyama[2], Xiang Li[3], Hironori Hojo[4], Ikuo Fujinaga[1], Takuya Katashima[1], Taku Saito[5], Ungil Chung[1], Naoyuki Sakumichi[1*] & Takamasa Sakai[1*]

[1]Department of Bioengineering, Graduate School of Engineering, The University of Tokyo, 7-3-1 Hongo, Bunkyo-ku, Tokyo 113-0033, Japan.

[2]Department of Materials Physics, Graduate School of Engineering, Nagoya University, Furo-cho, Chikusa, Nagoya 464-8603, Japan

[3]The Institute for Solid State Physics, The University of Tokyo, 5-1-5 Kashiwanoha, Kashiwa, Chiba 277-8581, Japan

[4]Center for Disease Biology and Integrative Medicine, Division of Clinical Biotechnology, School of Medicine, The University of Tokyo, 7-3-1 Hongo, Bunkyo-ku, Tokyo 113-0033, Japan

[5]Department of Orthopaedic Surgery, Faculty of Medicine, The University of Tokyo, 7-3-1 Hongo, Bunkyo-ku, Tokyo 113-0033, Japan.

*Corresponding authors.
Email: sakumichi@tetrapod.t.u-tokyo.ac.jp (N.S.), sakai@tetrapod.t.u-tokyo.ac.jp (T.S.)

**Abstract:**

Cosmic large-scale structures, animal flocks, and living tissues are non-equilibrium organized systems created by dissipative processes. Despite the uniqueness, the realization of dissipative structures is still difficult. Herein, we report that a network formation process in a dilute system is a dissipative process, leading to percolation induced gel–gel phase separation (GGPS) in a prominent miscible polymer–water system. The dilute system, which forms a monophase structure at the percolation threshold, eventually separates into two gel phases in a longer time scale as the network formation progresses. The dilute hydrogel with GGPS exhibits an unexpected mesoscale co-continuous structure and induces adipose growth in subcutaneous. The formation mechanism of GGPS and a cosmic large-scale structure is analogous, in terms of attractive interactions in a diluted system driving phase separation. This unique phenomenon unveils the possibility of dissipative structures enabling advanced functionalities and will stimulate research fields related to dissipative structures.

**Main Text:**

Dissipative processes occasionally form non-equilibrium organized structures that are difficult to manufacture. Our universe has a large-scale organized structure comprising voids and walls, which is formed in a non-equilibrium process[1,2]. In addition, membrane-less organized structures, such as nucleoli, stress granules, and Cajal bodies, are produced by liquid–liquid phase separation (LLPS) in dynamic living systems[3]. LLPS is associated with the development of several diseases, including amyotrophic lateral sclerosis, cataracts, neurodegenerative diseases, and aging[4].

Phase separation is extensively observed in living systems, partly because of the presence of macromolecules[5]. Compared to small-molecule systems, macromolecular systems have fewer molecules owing to their large molar masses. Moreover, mixing is not strongly favored in macromolecular systems because the mixing entropy depends on the number of molecules. The phase-separation tendency is further enhanced when macromolecules are connected to form a three-dimensional macromolecular network, that is, a gel. Phase separation induces macroscopic volume changes due to the solidity of gels. This has received significant research-based and application-based attention[6]. However, despite extensive research, the fundamentals of gels are not yet fully understood. Two substantial achievements have been recently realized in this regard, in terms of discoveries of the universal scaling law of osmotic pressure[7] and negative energy elasticity[8].

This study identified gel–gel phase separation (GGPS) occurring during gelation only at solute concentrations below a specific concentration (**Fig. 1**). GGPS was induced through a dissipative process and eventually arrested through gelation. The hydrogel with GGPS induced adipose growth in vivo, which has not been observed in analogous gels without phase separation.

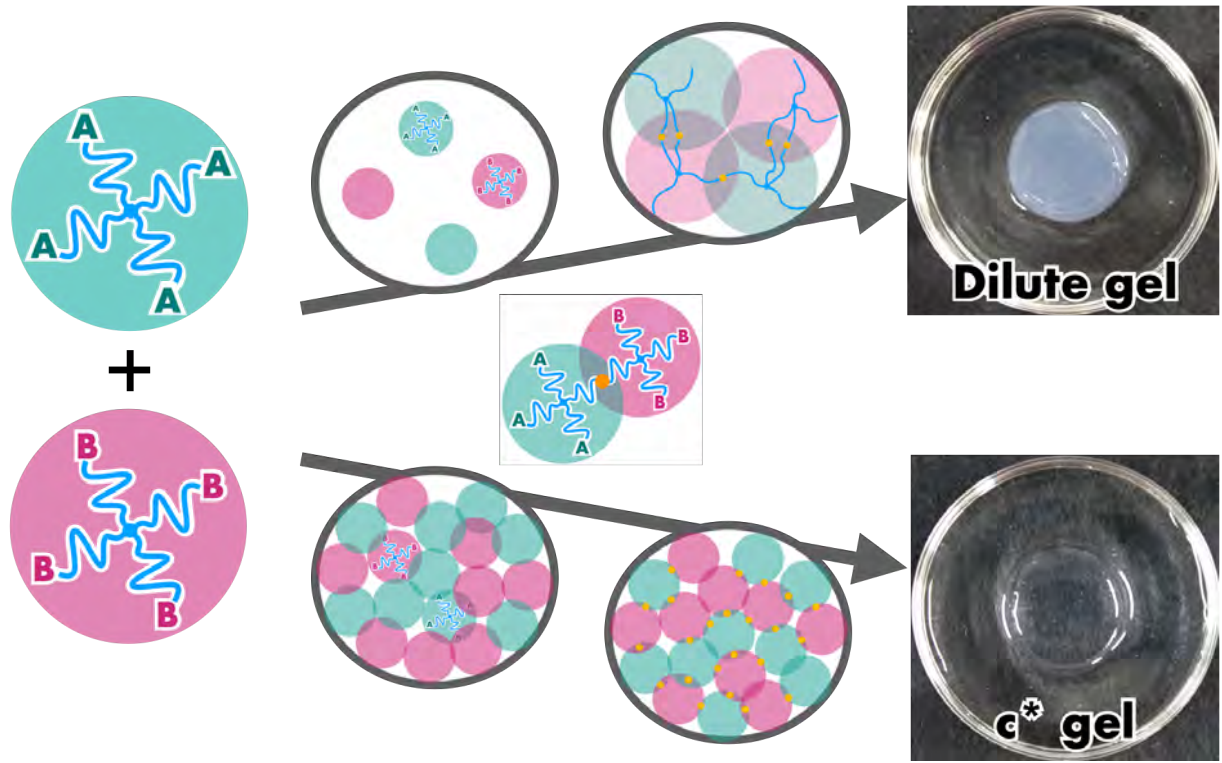

**Fig. 1 | Schematic of the gelation reaction for yielding the dilute and c\* gels.** Tetra-functional polymeric precursors with mutually reactive functional groups are coupled under aqueous conditions. The reactive groups, maleimide (A) and thiol (B), form covalent linkages. The dilute (turbid) and c\* (transparent) gels are formed at $c$ = 10 and 60 g/L, respectively. The precursors are mixed under stoichiometrically balanced conditions. In the formation process of $c^*$ gel (bottom), bond formation between the PEG precursors does not significantly influence the spatial arrangement of the PEG units. On the other hand, in the formation process of the dilute gel (upper), the spatial arrangement of the PEG chains is strongly influenced by the bond formation, which induces the expansion of polymeric clusters. This expansion occurs because of the agglomeration of clusters with fractal dimensions less than 3. The volume occupied by the clusters increases with gelation because of the expansion and encompasses the system prior to the gel point.

**Results**

**Abnormal clouding in a miscible system**

A water–polyethylene-glycol (PEG) binary system was investigated in this study. PEG is a prominent hydrophilic macromolecule whose aqueous solutions and hydrogels have been extensively investigated[9,10]. Hydrogels were synthesised in this study using mutually reactive tetra-functional PEG precursors and tuning the PEG concentration ($c$) in the range of 10–90 g/L (**Fig. 1**). This molecular design enables a high reaction conversion of up to 95%, regardless of the polymer concentration[11]. The PEG precursors were miscible with water in the investigated range, and transparent gels were formed above the overlap concentration, $c^*$. Here, $c^*$ ($\approx$ 60 g/L) is the concentration at which the territory occupied by hydrated PEG precursors engulfs the entire system[12].

PEG concentrations below $c^*$ result in the gelling solution becoming cloudy as the reaction proceeds (**Fig. 2a**). The cloudiness of the gels increases as the PEG concentration decreases (**Fig. 2b**), which contradicts conventional notions regarding phase separation occurring at solute concentrations above a specific value[5]. Therefore, two representative gels (dilute and $c^*$) formed at $c$ = 10 and 60 g/L, respectively, were examined to elucidate this behaviour.

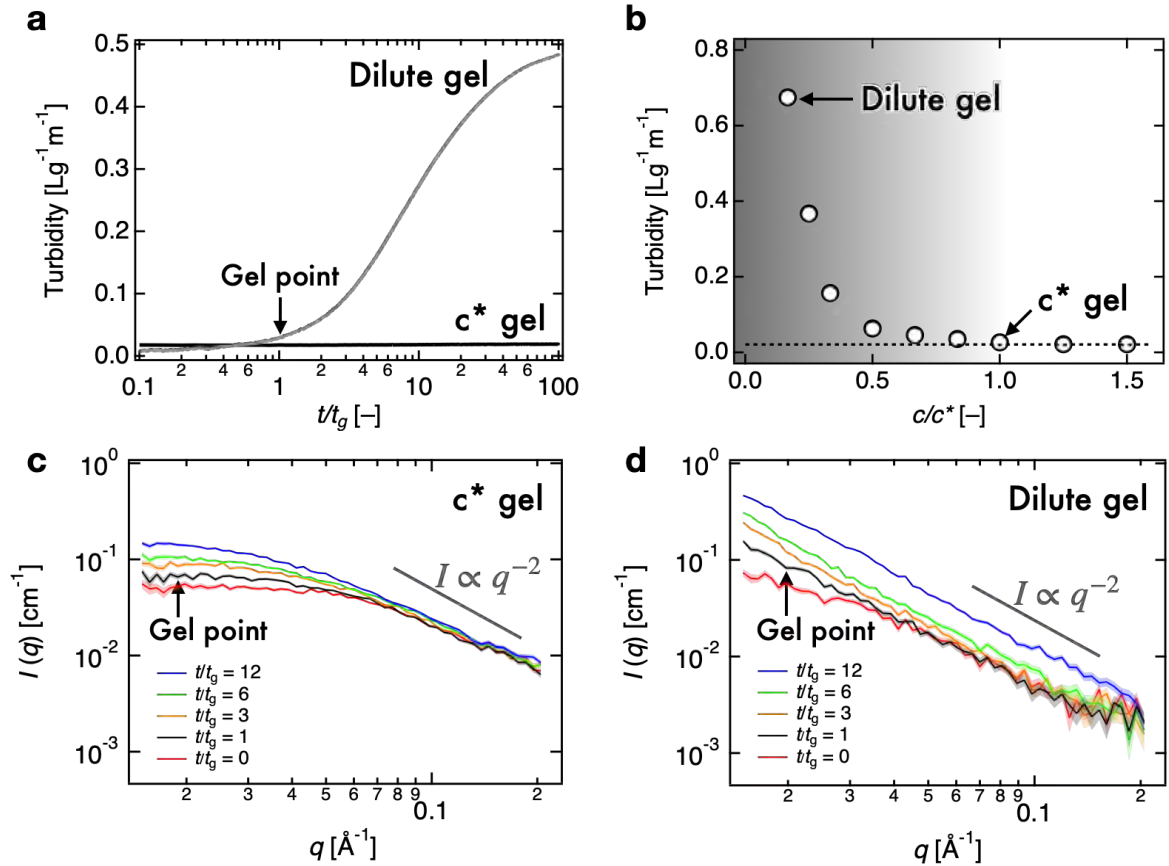

**Fig. 2 | Abnormal clouding in a miscible system. a,** Temporal changes in turbidity with gelation of the dilute (10 g/L; gray profile) and $c^*$ gels (60 g/L; black profile). The gelation time $t_g$ was used to normalize $t$. **b,** Equilibrium turbidity as a function of concentration normalized by the overlapping concentration. Time-resolved small-angle X-ray scattering (SAXS) profiles of the **c,** $c^*$, and **d,** dilute gels. The process of gelation was continuously monitored every 5 min. The various scattering curves (red to blue) correspond to the progression of gelation, with the black profiles representing the gel point (i.e., the sol–gel transition point).

## Structure formation during gelation

Time-resolved SAXS experiments were performed for the gelling solutions to exclude the possibility of the enhanced cloudiness being caused by precipitation, which occasionally competes with gelation[13,14]. However, the results did not provide any indication of precipitation being involved. The scattering profiles did not conform to the fractal relationship $I \propto q^{-4}$, which originates from the sharp interface formed by precipitation. Detailed analysis of the scattering profiles revealed the structural differences between the $c^*$ and dilute gels, in terms of the dependence of the scattering vector ($q$) on the scattering intensity ($I$).

The scattering profiles corresponding to the $c^*$ gel change slightly during gelation (**Fig. 2c**), which indicates that bond formation between the PEG precursors does not significantly influence the spatial arrangement of the PEG units. The molecular representation corresponding to $c > c^*$ (**Fig. 1**; bottom schematic) clarifies this observation, based on the polymer precursors encompassing the pre-gel solution, which leaves little room for rearrangement[14]. A further increase in precursor concentration ($c \approx 2c^*$) was found[15] to result in a complete overlap of the scattering profiles before and after the gel point.

In contrast, the scattering profiles of the dilute gel exhibit drastic changes during gelation (**Fig. 2d**). Initially, the scattering profiles conform to the fractal relationship $I \propto q^{-2}$, which originates from the internal correlation of the precursors with mass fractal dimension $D = 2$, followed by a crossover to a plateau, $I \propto q^0$, corresponding to the size of the precursors, $R$ ($q \sim 1/R \sim 0.05$ Å$^{-1}$)[16]. $D$ is a measure of the space occupied by the fractal object and is defined as $L \sim M^{1/D}$, where $L$ is the representative length scale and $M$ is the mass of the object. As the gelation proceeds, the fractal region expands to accommodate a considerably lower $q$ (~0.02 Å$^{-1}$), indicating the growth of polymeric clusters. The invariance of the internal correlation with $D = 2$ suggests a homogeneous distribution of the polymer segment, even at the gel point. Given that the original precursors do not engulf the system, the homogeneous distribution of polymer segments is somewhat puzzling. The currently acknowledged gelation model[12] suggests that a homogeneous gel is formed at the gel point, even in a dilute system (**Fig. 1**; top schematic), because of the expansion of the polymeric clusters during gelation. This expansion occurs because of the agglomeration of clusters with fractal dimension less than 3. When two fractal clusters of mass $m_0$ at end-to-end distances of $l_0$ agglomerate to form a cluster at the end-to-end distance of $l_1$, the mass of the resulting cluster is the sum of the masses of the original clusters ($2m_0$). In contrast, assuming that the fractal dimension is maintained, the volume occupied by the clusters changes from $v_0 = 2l_0^3 = 2m_0^{3/D}$ to $v_1 = l_1^3 = (2m_0)^{3/D}$. The relationship $v_0 < v_1$ holds true only when the fractal dimension of the cluster is less than 3. The volume occupied by the clusters increases with gelation because of the expansion and encompasses the system prior to the gel point. The gel envelops the system as long as it is in the gel form. During the gelation of the dilute gel, the clusters grow to a mesoscale level (~$10^{-5}$ m)[17], leading to the existence of a mesoscale concentration fluctuation in the polymer network.

A unique change occurs in the scattering profiles of the dilute gel after the gel point. The scattering intensities continuously increase with time over the entire SAXS $q$-range, in marked contrast with the behavior of the $c^*$ gel. The increase in scattering intensity typically indicates an enhanced spatial correlation at a specified size scale, such as the appearance of Bragg peaks during crystallization of a polymer melt[18]. Similarly, the aforementioned observation suggests an increase in the spatial correlation over a wide size range. This change can presumably be attributed to the local condensation of the polymer chains. Notably, the fractal relationship, $I \sim q^{-2}$, is conserved throughout the process, indicating that the mass fractal dimension $D$ remains constant during condensation. These results suggest that the gel network mesoscopically phase-separates into the concentrated and dilute phases on a size scale larger than the range of the SAXS measurements (>100 nm).

**Molecular dynamics (MD) simulation**

The mesoscale structures were examined by performing coarse-grained MD simulations. Snapshots of the structures during gelation are shown in **Fig. 3**. The concentrated-gel scenario (**Fig. 3a**) results in insignificant concentration fluctuations. In contrast, the dilute condition

(**Fig. 3b**) produces increasing concentration fluctuations with advancing gelation, eventually forming a mesoscale phase-separated structure, similar to that observed experimentally. A clear increase in scattering intensity occurs under dilute conditions (**Supplementary fig. 1**), and the $q$-dependence of the structure factors $S(q)$ ($\sim q^{-2}$) appears to be consistent with the experimental SAXS data. The characteristic size of the porous structure is significantly larger than that of the precursors and does not increase with time. This feature is different from that of microscopic[19] or macroscopic phase separation[20], suggesting that mesoscopic phase separation occurs by a new mechanism related to gelation.

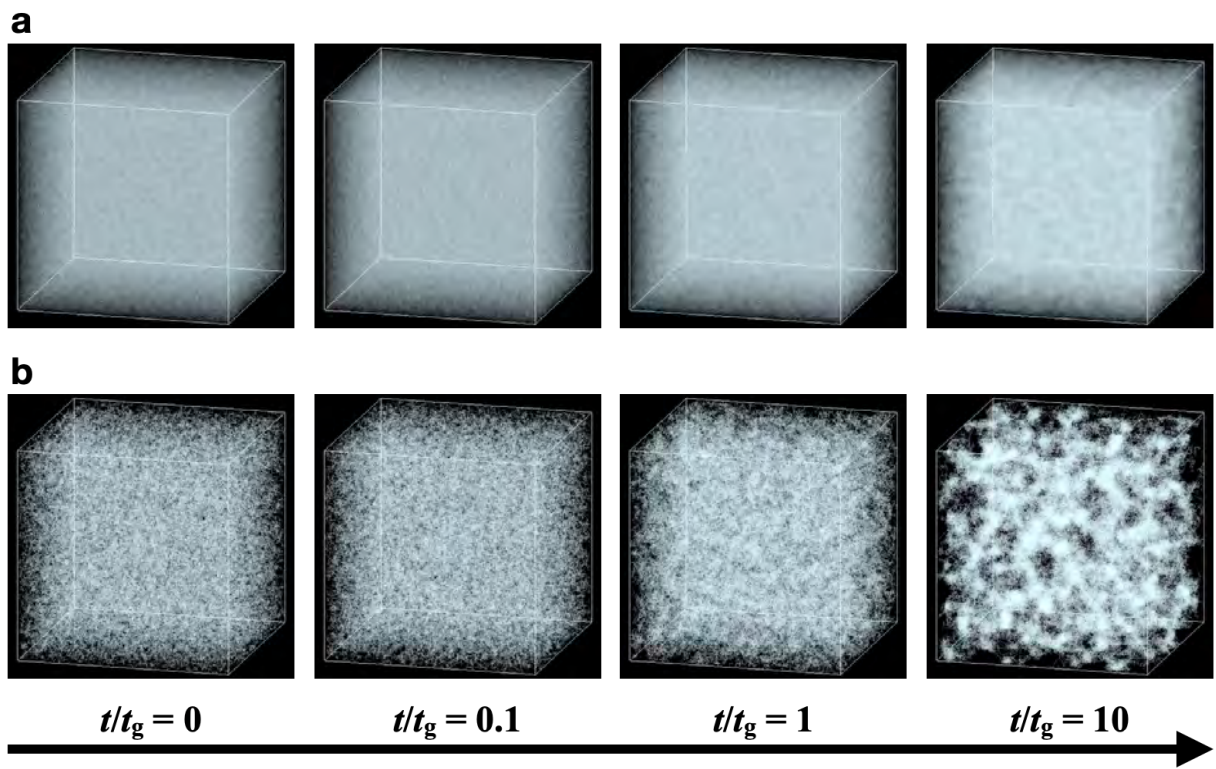

**Fig. 3 |** MD simulation results of network structures during gelation of **a,** concentrated ($c$ = 1.5, see Methods) and **b,** dilute gels ($c$ = 0.094). The values represent durations relative to the gelation time $t_g$. The structure factors $S(q)$ were calculated using the simulations to examine the correspondence with the experiments (**Supplementary fig. 1**). The time dependence of $S(q)$ indicates that the mesoscopic structures start growing at $t \cong t_g$.

*Equilibrium of an open system*

Gels exhibit two equilibria corresponding to closed and open systems[21]. The equilibrium of a closed system is achieved throughout the process of gelation, wherein the precursors react with each other to achieve a stable condition for forming an equilibrated polymer network. Equilibrium in an open system is achieved when the gel is immersed in a solvent, with the solvent molecules moving in and out of the gel to achieve equilibrium.

The equilibrium of an open system is discussed herein. The immersion of the gels in water results in swelling and deswelling of the *c\** and dilute gels, respectively (**Supplementary fig. 2**). The equilibrium state of an open system is determined by the balance between the osmotic and elastic pressures, which promote swelling and deswelling, respectively. Most gels swell like *c\** gel[7] because their osmotic pressure exceeds the elastic pressure, indicating that the deswelling observed in the dilute gel is abnormal. The cloudiness of the dilute gel was further enhanced by deswelling, and a white turbid gel was obtained, as shown in **Fig. 1**.

**Examination of GGPS**

The enhanced cloudiness suggests the formation of a clear mesoscale structure in the dilute gel. **Fig. 4a** indicates that the *c\** gel does not present any structure until day 7. This finding is reasonable because the mesh size of the polymer network is conventionally on the order of $10^{-9}$ m[21] and cannot be microscopically observed. However, a clear co-continuous mesoscale network structure appears in the dilute gel, as shown in the day-4 and day-7 images, although no structure is present on day 0. The featureless day-0 pattern is consistent with the mesoscopic phase separation indicated by SAXS.

Three features are visible in the day-7 mesoscale structure of the dilute gel. First, the size of the structure is on the order of $10^{-5}$ m, which is $10^4$ times larger than the conventional mesh size. Second, the dilute phase fluoresces, indicating the presence of PEG; therefore, GGPS occurs. Third, the volume ratio of the two phases is clearly uneven, whereas that of the conventional co-continuous phase separation is even[22]. The fact that phase separation occurs at concentrations below $c^*$ to yield a homogeneous gel must be re-emphasized. This strongly suggests that phase separation originates from "gelation below $c^*$".

**Supplementary fig. 3** indicates the temperature-based stability of the mesoscale structure in a range of 4–60 °C under atmospheric pressure. A further increase in temperature to 90 °C results in the shrinking and increased transparency of the dilute gel and disappearance of the mesoscale structure. Shrinkage at elevated temperatures has been observed in conventional PEG hydrogels and reflects the hydrophobic nature of PEG at high temperatures[23]. However, the reversal of the temperature from 90 °C to 25 °C results in the dilute gel becoming cloudy and the appearance of a similar mesoscale structure (**Supplementary fig. 3**).

The GGPS observed in this study leads to an interesting anomalous state that violates the (Gibbs) phase rule, $F = C - P + 2$, in thermodynamic equilibrium. Here, $F$, $C$, and $P$ are the numbers of thermodynamical degrees of freedom, components, and phases, respectively. The dilute gel immersed in water is a PEG–water two-component system ($C = 2$) with three phases (concentrated and dilute phases, and water; $P = 3$). Thus, the phase rule results in $F = 1$, that is, the three-phase coexistence is allowed only at a specific temperature under atmospheric pressure[20]. However, the three phases coexist across a wide temperature range, apparently violating the phase rule. Notably, states that violate the phase rule can exist when the intrinsic length scale acts as an additional field variable. In fact, model calculations using numerical simulations have predicted the occurrences of a four-phase coexistence in a one-component system[24] and a five-phase coexistence in a two-component system[25]. In these cases, the generalized phase rule[25], $F = C - P + S + 2$, is satisfied, where $S$ is the number of intrinsic length scales. Thus, the dilute gel experimentally achieves a thermal equilibrium state that violates the phase rule, which may be justified by the existence of an intrinsic length scale (the mesh size).

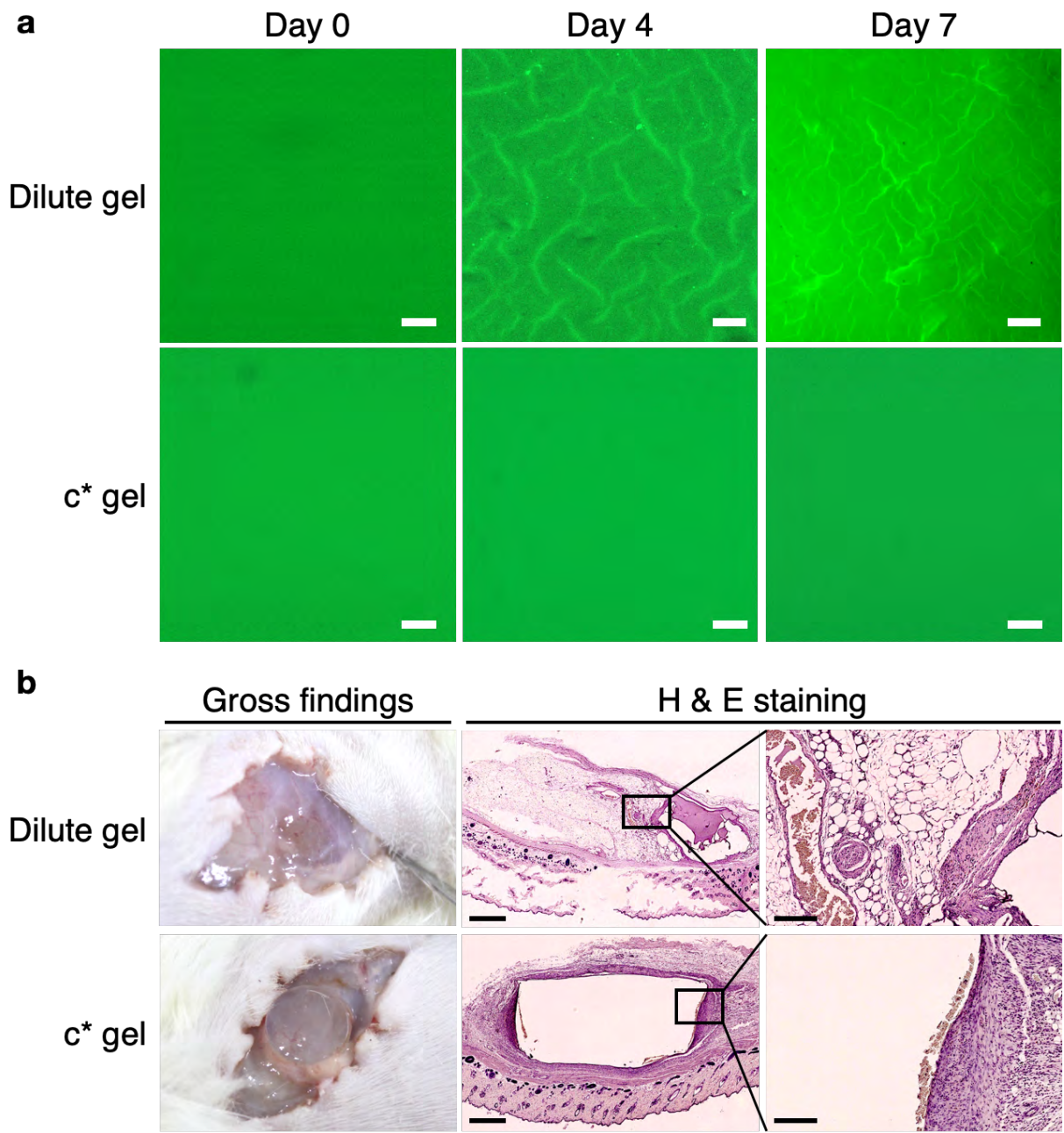

**Fig. 4 | Mesoscale structure and interactions with living tissue. a,** Confocal laser scanning microscopy images of the dilute and *c*\* gels (scale bars: 100 μm). **b,** Gross and histological observations of the dilute and *c*\* gels (the scale bars in the center and right photos represent 1 mm and 250 μm, respectively). The dilute and *c*\* gels were subcutaneously implanted on the

backs of rats to investigate gel interactions with living systems and observed two weeks after implantation.

**Mechanism of GGPS**

GGPS is spontaneously induced by gelation in the homogeneous polymer–water system. Here, we explain the mechanism of GGPS chronologically. The dilute system expands due to cluster agglomeration and exhibits mesoscale concentration fluctuations at the gel point[26]. The macroscopic stiffness is infinitesimally small just above the gel point and is therefore exerted by a small fraction of the PEG chain. The system presumably has a mesoscale heterogeneous distribution of elasticity with a size corresponding to that of the concentration fluctuation at the gel point. In the mesoscale area with high elasticity, elasticity works as an attractive force and condenses the adjacent PEG units. Notably, the system is far from equilibrium, and an additional crosslinking reaction occurs, because the reaction conversion at the gel point is only up to 60% at $c$ = 10 g/L[12]. The crosslinking reaction takes place efficiently in the condensed area and further enhances the local elasticity. This positive feedback shifts the equilibrium from the original one, which involves a dissipative process driven by the chemical reaction. The dissipative process induces mesoscale soft GGPS as revealed by SAXS and the MD simulations. Notably, the dissipative structure was arrested and frozen permanently during gelation.

In an open environment that enables macroscopic volume changes, further local condensation occurs and forms a distinct interface. The dense phase becomes even denser to minimize the interfacial energy, further driving GGPS. Consequently, an uneven co-continuous structure is formed via GGPS in a homogeneous miscible system.

**Interaction with a living system**

The mesoscale structure of the dilute gel evokes the extracellular matrix in living systems. Therefore, the interactions of the dilute and $c$* gels with living systems were investigated (**Fig.**

4b, Supplementary fig. 4). Surprisingly, only the dilute gel dissolves in vivo and is substituted into an adipose-like tissue, including the vascular structure. This behavior is distinct from that of the *c*\* gel; no cell invasion or degradation is apparent, which is consistent with similar experiments conducted on hydrogels with a PEG backbone[27]. To the best of our knowledge, this is the first observation of the PEG-hydrogel-based cell invasion and tissue substitution without any specific bioactive motifs, such as cell adhesion peptides (RGD sequences) or apparent degradability. It bears repeating that the cell-invading dilute gel and cell-inert *c*\* gel differ only in terms of the initial PEG concentration. Moreover, the unique mesoscale structure formed via GGPS provides an unexpected bio-affinity.

**Discussion**

The results reported in this manuscript unveil the unique properties of dissipative structures that have not been predicted by their chemical structures. GGPS is presumably related to the formation of structures in human bodies because they are analogues to turbid hydrogels. Furthermore, similarities between GGPS and the formation of the universe were noted. The concentration fluctuation grows to the mesoscale at an early stage (gelation and inflation) and develops into a large-scale phase-separated structure assisted by an attractive interaction (elasticity and gravity) far from equilibrium. Therefore, the synthesis of this dilute gel mimics the creation of the universe on an elementary level. These findings are anticipated to stimulate discussions in materials science, biology, cosmology, and fields related to dissipative structures.

**Methods**

**Preparation of PEG hydrogels**

Four-armed poly(ethylene glycol)s ($Mw$ = 10,000) functionalized with sulfhydryl (tetra-PEG-SH) and maleimide (tetra-PEG-MA) were purchased from NOF Co. Ltd. (Tokyo, Japan). The PEG hydrogels were prepared by dissolving fixed amounts of tetra-PEG-SH and tetra-PEG-MA in citrate phosphate buffer (50 mM, pH 4.2). Equal amounts of these precursors were subsequently mixed and immediately poured into a mold. The gelation reaction was conducted at 25 °C for 24 h prior to the experiments. The overlapping concentration $c^*$ of the PEG precursors was approximately 60 g/L, and the hydrogels prepared at 10 g/L and 60 g/L were referred to as dilute and $c^*$ gels, respectively. Hydrogels prepared using the following concentrations were investigated and characterized based on the normalized concentration $c/c^*$: 10, 15, 20, 30, 40, 50, 60, 75 and 90 g/L ($c/c^*$ = 0.16, 0.25, 0.33, 0.5, 0.66, 0.83, 1.0, 1.25 and 1.5, respectively).

**UV–vis spectra**

The precursors of the samples were mixed and poured into plastic cells with an optical length of 10 mm. The transmittance at λ = 400 nm was measured using a UV–vis spectrophotometer (V-670, JASCO Corp., Tokyo, Japan) at 25 °C every 5 s for 48 h. The turbidity was estimated by normalizing the absorbance with the polymer concentration (g/L).

**Optical observation of hydrogels**

The precursors were mixed and poured into Teflon molds (diameter: 15 mm, height: 7 mm) and allowed to sit for 24 h at 25 °C to enable gelation. The gel samples were carefully removed from the molds and subsequently immersed in distilled water at 25 °C. Optical imaging and turbidity analysis of the hydrogels were performed by macroscopic observations of the gel samples on days 0, 2, 4, and 7. The relative turbidity was estimated based on the background intensity $T_b$ and hydrogel intensity $T_h$ calculated using the Image J system.

**SAXS**

SAXS measurements were carried out at the BL-6A beamline of the Photon Factory, High Energy Accelerator Research Organization, KEK (Ibaraki, Japan). An incident X-ray wavelength of 0.15 nm, sample-position beam diameter of ~0.25 × 0.50 mm$^2$, exposure time per sample of 30 s, and sample-to-detector distance of 2.54 m were employed. The scattering profiles were collected using a 2D hybrid pixel detector (PILATUS3 1 M, DECTRIS Ltd., Switzerland). All measurements were conducted at ambient temperature (~25 °C). A 1-mm-thick gel sample was placed in a custom-made planar cell and sealed with two 30-μm-thick glass windows. The scattered intensities were circularly averaged to obtain 1D intensity profiles and subsequently corrected for the incident beam flux, sample absorption, sample thickness, exposure time, and cell and solvent scattering using a custom-made data reduction package (Red2D; https://github. com/hurxl/Red2D) within a scientific data-analysis-software package (Igor Pro 8, WaveMetrics). The intensity was plotted as a function of the magnitude of the scattering vector $q$. Glassy carbon (National Institute of Standards and Technology (NIST), USA) and silver behenate (Nagara Science, Japan) standards were used for calibration. All measurements were performed at 25 °C.

**Confocal laser scanning microscopy**

CLSM images were obtained to visualize the mesoscale phase separation of the hydrogels. Fluorescence-labelled PEG was first prepared to facilitate CLSM. Briefly, 1,000 mg of tetra-PEG-SH was dissolved in 20 mL of distilled water and stirred for 10 min at ambient temperature. Separately, 1 mg of Alexa Fluor™ 488 C5 maleimide (Thermo Fisher Scientific,

USA) was dissolved in 1 mL of dimethyl sulfoxide, and the solution of 128.6 µL (0.0125 eq. vs. tetra-PEG-SH) was added to the tetra-PEG-SH solution. The mixture was allowed to sit for 3 h at ambient temperature and subsequently dialyzed against distilled water for 3 h to remove unreacted molecules and freeze-dried to obtain tetra-PEG-SH functionalized partly with Alexa Fluor 488 as a faint yellow powder (yield: 950 mg).

Gel precursors prepared using fluorescently labelled tetra-PEG-SH and tetra-PEG-MA were poured into a Teflon mold (diameter: 15 mm, height: 7 mm), and allowed to sit for 24 h at 25 °C to enable gelation. The prepared hydrogels were carefully removed from the mold and immersed in distilled water at 25 °C for 7 d. The CLSM images were obtained using an LSM 800 setup (ZEISS, Germany).

**Temperature analysis of the gels**

Variations in the mesoscale phase separation with temperature were examined by immersing the hydrogels in distilled water for 7 d and exposing them to various temperatures (4, 15, 40, 60, and 90 °C). Optical and CLSM imaging as well as turbidity analysis were performed as described above after 24 h at a specific temperature.

**MD simulation**

Coarse-grained MD simulations were performed to examine the phase-separation behaviors during and after gelation. A single tetra-PEG molecule was modelled using five particles: one junction point and four centers of mass (CMs) of the sub-chains. Each CM was connected to the junction point by a harmonic potential (a linear spring): $u_{\text{harmonic}}(r) = (k_0/2)\,r^2$ ($k_0$: spring constant). The CMs repelled each other because of the steric repulsion potential: $u_{\text{steric}}(r) = \varepsilon_0/(2\pi q_0)^{3/2} \exp(-r^2/2q_0^2)$ ($\varepsilon_0$ and $q_0$ represent the repulsive interaction strength and characteristic size, respectively). For simplicity, the steric repulsion between the different molecules was neglected. The system consisted of equal numbers of two types of tetra-PEG molecules. The two CMs of the different types of tetra-PEG molecules reacted with a constant probability to form a crosslink.

Initially, the equilibrated molecules without any crosslinks were placed into a periodic box. The positions of the particles were subsequently evolved using the Langevin equation (the Brownian dynamics scheme), resulting in the crosslinking reactions occurring stochastically. After the reaction, two CMs were connected by another harmonic potential. Dimensionless units were employed based on the characteristic size of the sub-chain, characteristic diffusion time of the sub-chain, and thermal energy becoming unity. The details of the simulation model and numerical scheme are provided elsewhere. $k_0 = q_0 = 1$ and $\varepsilon_0 = 40$ were used as the parameters for the potentials, with a reaction rate, $k_{\text{reaction}}$, of 0.1. The system size was set to $128^3$ and the time step to $\Delta t = 0.01$. The total number of molecules in the system was altered, and the concentration of the system ranged from $c = 0.0474$ to 6. The simulations were performed for a sufficiently long time until most CMs had been crosslinked.

Clusters connected by crosslinks were extracted from the simulation results, and the gelation time $t_g$ was estimated as the time at which the system was almost covered by large clusters. This condition can be roughly interpreted as the percolation point. The structures were visualized by drawing bonds connecting the particles. Moreover, the structure factors $S(q)$ were calculated by mapping the junction points to density fields, which were subsequently subjected to the Fourier transform.

**In vivo study on rats**

All animal experiments were performed according to the protocol approved by the Animal Care and Use Committee of the University of Tokyo. Twelve-week-old female Wistar rats weighing 170–200 g were used in this study. The rats were randomly divided into two groups, with one being subjected to subcutaneous implantation and injection of the *c*\* gel (*n* = 5), and the other receiving the dilute gel (*n* = 5) through identical procedures. The rats were housed and provided laboratory rat chow and water ad libitum and exposed to a 12 h light/dark cycle at a room temperature (22 °C).

**Subcutaneous injection of hydrogels**

All surgical procedures were performed under general anesthesia and sterile conditions. For general anesthesia, a combination anesthetic composed of 0.3 mg/kg medetomidine, 4.0 mg/kg midazolam, and 5.0 mg/kg butorphanol was intraperitoneally administered to each rat at 0.5 mL/100 g of body weight. Epilation of the skin at the surgical site was performed prior to the incision. A 4-cm-long skin incision was made in the epilated back right side, and the subcutaneous tissue was exposed. After creating space for the hydrogel by separating subcutaneous tissue from the fascia, disk-shaped hydrogels (diameter: 10 mm, height: 2 mm) were gently implanted in each group. After the surgical procedure, the skin was closed with skin staplers and 5 mg/kg enrofloxacin was intramuscularly administered for antibiotic prophylaxis. The rats were sacrificed two weeks after the procedure and subjected to histological analysis.

**Data availability**

The data that support the findings of this study are available from T.S. upon reasonable request.

**Acknowledgements**

This study was supported by the Japan Society for the Promotion of Science (JSPS) through Grants-in-Aid for JSPS Research Fellow Grant Number 20J01344 to S.I., Early Career Scientists grant number 19K14672 to N.S., Scientific Research (B) grant number 18H02027 to T.S., and Transformative Research Areas grant number 20H05733 to T.S. This study was also supported by the Japan Science and Technology Agency (JST) PRESTO Grant Number JPMJPR1992 to T.U., and JST CREST Grant number JPMJCR1992 to T.S.

**Author contributions**

All authors contributed to writing the manuscript and discussing the results. Experiments were carried out and planned, and analysis was performed by S. I., I. F., X. L., and U. C. Discussions regarding the mechanism of GGPS involved S. I., N. S., T. K., and T.S. Animal experiments were performed by S. I., Y. I., and T. S. T.S. conceived the idea and supervised the project.

**Competing interests**

The authors declare no competing interests.

# Supplementary information

## Percolation induced gel–gel phase separation in a dilute system

Shohei Ishikawa[1], Ikuo Fujinaga[1], Xiang Li[2], Yasuhide Iwanaga[3], Takashi Uneyama[4], Naoyuki Sakumichi[1], Takuya Katashima[1], Taku Saito[3], Ungil Chung[1] & Takamasa Sakai[1]*

[1]Department of Bioengineering, Graduate School of Engineering, The University of Tokyo, 7-3-1 Hongo, Bunkyo-ku, Tokyo, Japan.

[2]The Institute for Solid State Physics, The University of Tokyo, 5-1-5 Kashiwanoha, Kashiwa, Chiba 277-8581, Japan

[3]The Department of Orthopaedic Surgery, Faculty of Medicine, The University of Tokyo, 7-3-1 Hongo, Bunkyo-ku, Tokyo, Japan.

[4]Department of Materials Physics, Graduate School of Engineering, Nagoya University, Furo-cho, Chikusa, Nagoya 464-8603, Japan

*Corresponding author. Email: sakai@tetrapod.t.u-tokyo.ac.jp (T.S.)

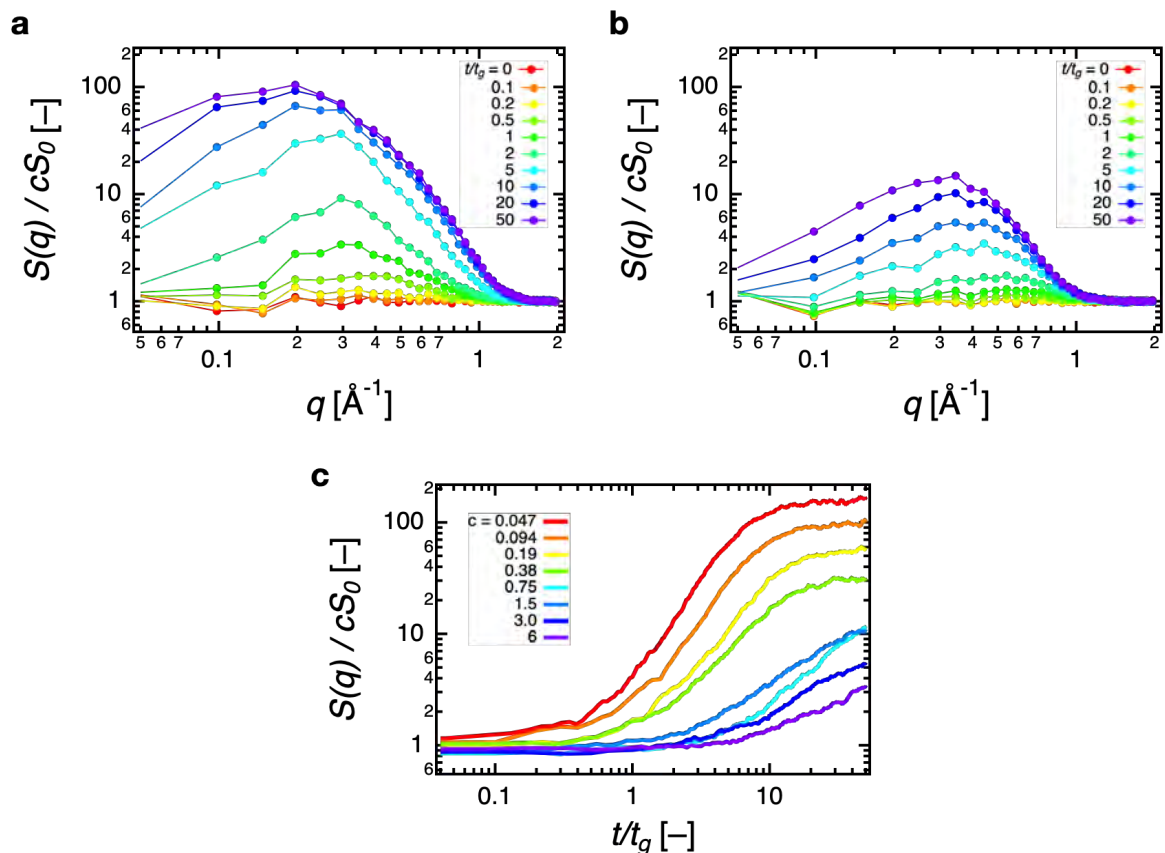

**Supplementary fig. 1 | Structure factors of gels calculated using the simulation snapshots.** Structure factors $S(q)$ for the **a**, dilute ($c = 0.094$) and **b**, concentrated gels ($c = 1.5$). The structure factors $S(q)$ were normalized by the concentration $c$ and the initial structure factors $S_0$ to ensure unity of the normalized structure factors of the homogeneous systems (initial structures). **c**, Growth behavior of normalized structure factors for $q = 0.2$ corresponding to gels prepared at different concentrations.

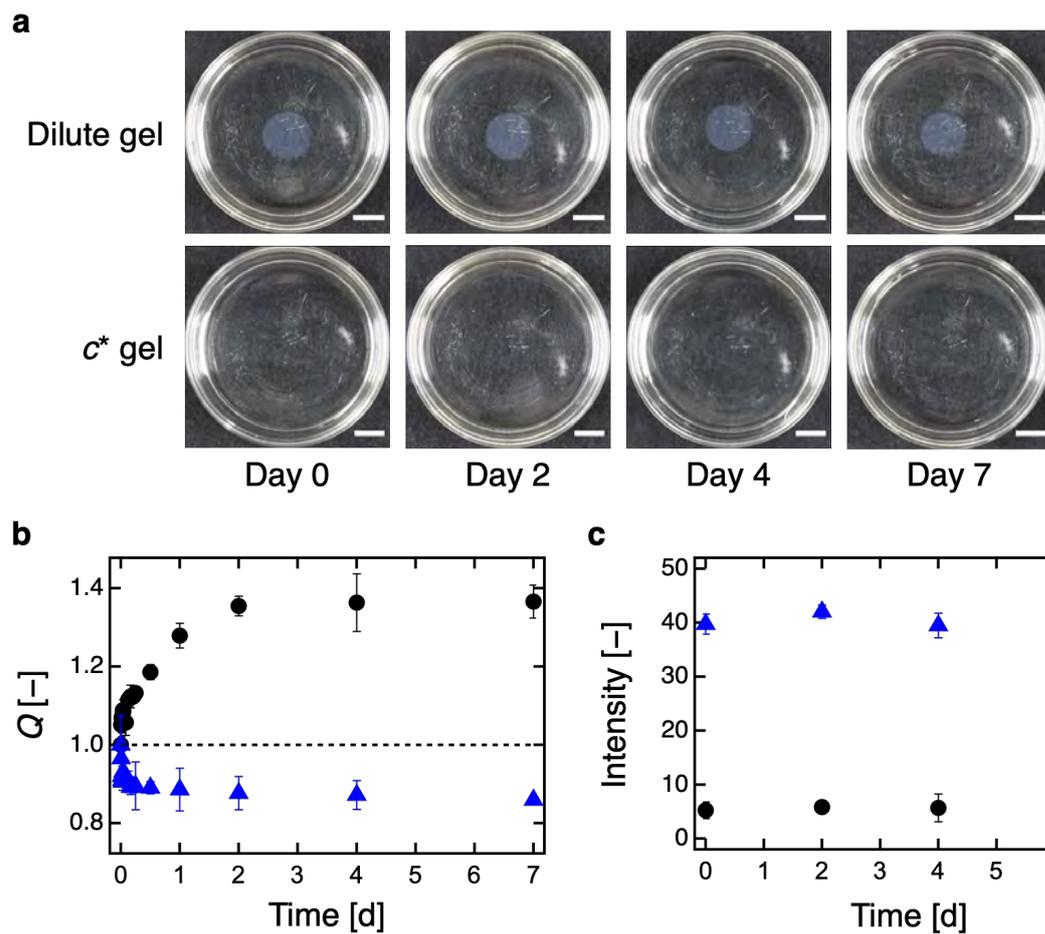

**Supplementary fig. 2 | Phase separation as an open condition. a,** Photos of the dilute and $c^*$ gels upon immersion in distilled water for 0, 2, 4, and 7 d (scale bars: 10 mm). **b,** Time-evolution of the swelling/deswelling and **c,** turbidity of the dilute (blue) and $c^*$ gels (black).

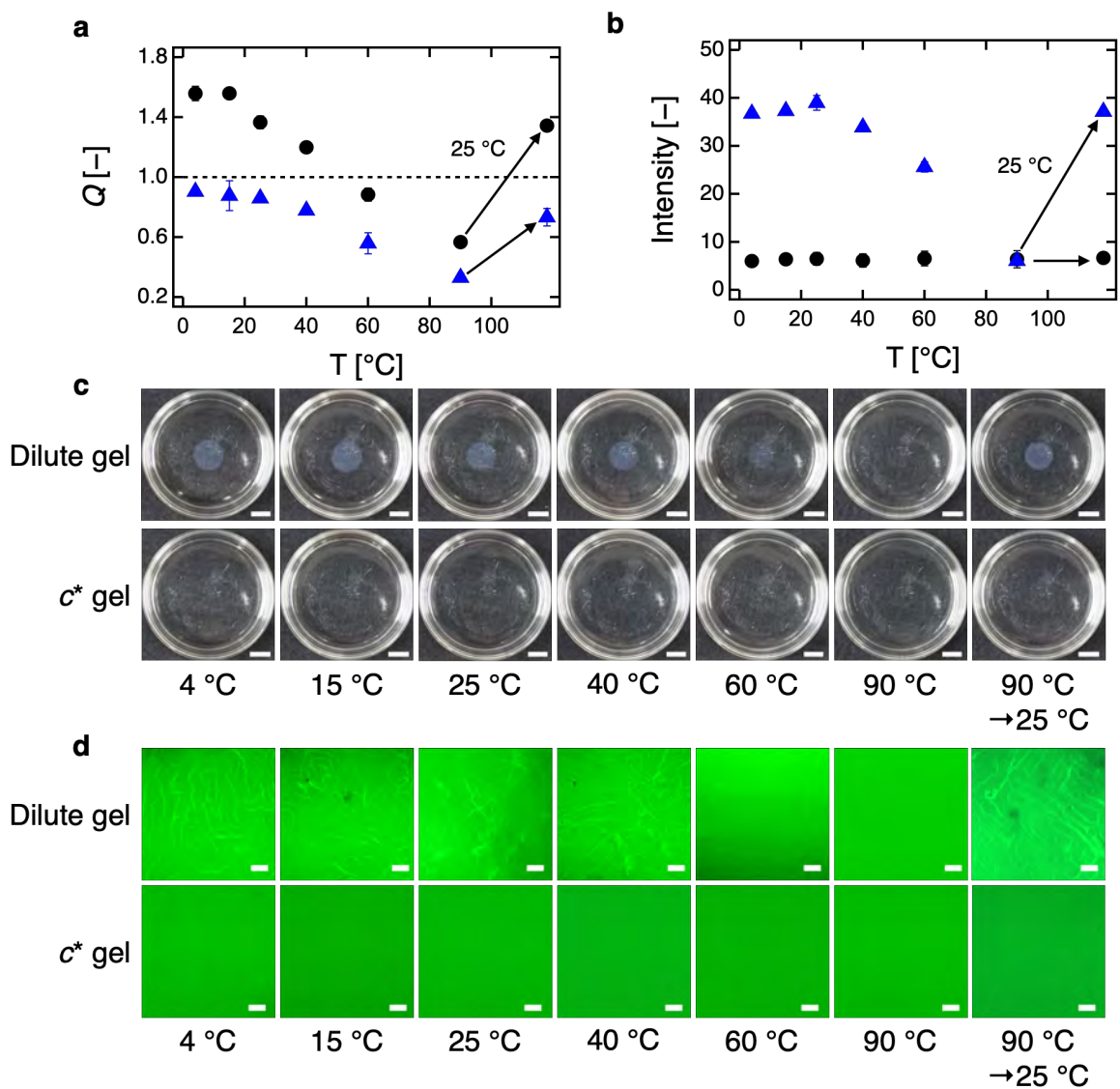

**Supplementary fig. 3 | Thermodynamic stability of the mesoscale structure. a,** Swelling/deswelling and **b,** turbidity of the dilute (blue) and $c^*$ gels (black). The hydrogels were immersed in distilled water for 7 d, exposed to various temperatures for 1 d, and analyzed using the ImageJ software. **c,** Photos of the $c^*$ and dilute gels exposed to various temperatures (scale bars: 10 mm). D, Fluorescence microscopy images of the dilute and $c^*$ gels (scale bars: 100 μm).

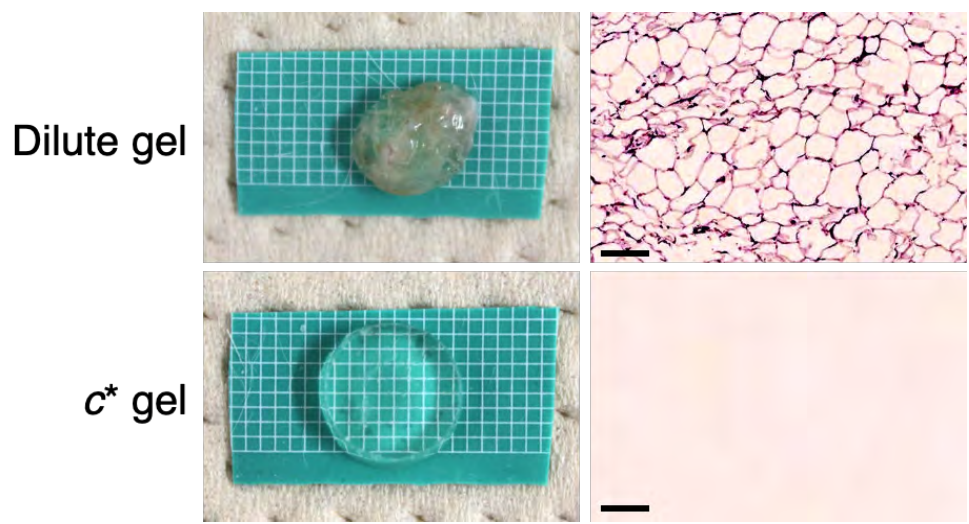

**Supplementary fig. 4 | Gross and histological findings with the dilute and *c\** gels (scale bars: 100 μm).**